\documentstyle[12pt,aaspp4]{article}

\begin{document}

\title{Complete RXTE Spectral Observations of the Black Hole X-ray Nova XTE~J1550--564}

\author{Gregory J. Sobczak\altaffilmark{1}, Jeffrey E. McClintock\altaffilmark{2},
Ronald A. Remillard\altaffilmark{3}, Wei Cui\altaffilmark{3}, Alan M.
Levine\altaffilmark{3}, Edward H.  Morgan\altaffilmark{3}, Jerome A. 
Orosz\altaffilmark{4}, and Charles D. Bailyn\altaffilmark{5}}

\altaffiltext{1}{Harvard University, Astronomy Dept., 60 Garden St. MS-10, Cambridge, MA
02138; gsobczak@cfa.harvard.edu}
\altaffiltext{2}{Harvard-Smithsonian Center for Astrophysics, 60 Garden St. MS-3, Cambridge, MA
02138; jem@cfa.harvard.edu}
\altaffiltext{3}{Center for Space Research, MIT, Cambridge, MA 02139; rr@space.mit.edu,
cui@space.mit.edu, aml@space.mit.edu, ehm@space.mit.edu}
\altaffiltext{4}{Sterrekundig Instituut, Universiteit Utrecht, Postbus 80.000, 3508 TA Utrecht,
The Netherlands; J.A.Orosz@astro.uu.nl}
\altaffiltext{5}{Dept. of Astronomy, Yale University, P. O. Box 208101, New Haven, CT
06520; bailyn@astro.yale.edu}

\begin{abstract} 

We report on the X-ray spectral behavior of XTE~J1550--564 during its 1998-99
outburst.  XTE~J1550--564 is an exceptionally bright X-ray nova and is also the third
Galactic black hole candidate known to exhibit quasiperiodic X-ray oscillations above
50~Hz.  Our study is based on 209 pointed observations using the PCA and HEXTE
instruments onboard the {\it Rossi X-ray Timing Explorer} spanning 250 days and
covering the entire double-peaked eruption that occurred from 1998 September until
1999 May.  The spectra are fit to a model including multicolor blackbody disk and
power-law components.  The spectra from the first half of the outburst are dominated
by the power-law component, whereas the spectra from the second half are dominated by
the disk component.  The source is observed in the very high and high/soft outburst
states of black hole X-ray novae.  During the very high state, when the power-law
component dominated the spectrum, the inner disk radius is observed to vary by more
than an order of magnitude; the radius decreased by a factor of 16 in one day during a
6.8~Crab flare.  If the larger of these observed radii is taken to be the last stable
orbit, then the smaller observed radius would imply that the inner edge of the disk is
inside the event horizon!  However, we conclude that the apparent variations of the
inner disk radius observed during periods of increased power-law emission are probably
caused by the failure of the multicolor disk/power-law model; the actual physical
radius of the inner disk may remain fairly constant.  This interpretation is supported
by the fact that the observed inner disk radius remains approximately constant over
120 days in the high state, when the power-law component is weak, even though the disk
flux and total flux vary by an order of magnitude.  The mass of the black hole
inferred by equating the approximately constant inner disk radius observed in the
high/soft state with the last stable orbit for a Schwarzschild black hole 
is $M_{BH} = 7.4~M_{\odot}(D/6~kpc) (\cos i)^{-1/2}$.  

\end{abstract}


\section{Introduction}

The X-ray nova and black hole candidate XTE~J1550--564 was discovered with the All Sky
Monitor (ASM; Levine et al.~1996) onboard the {\it Rossi X-ray Timing Explorer} (RXTE)
just after the outburst began on 1998 September~6 (Smith et al.~1998).  The source
exhibited a flare on 1998 September~19-20 that reached 6.8~Crab (or 1.6~$\times
10^{-7}$~erg~s$^{-1}$~cm$^{-2}$) at 2--10~keV.  The discovery of XTE~J1550--564
prompted a series of almost daily pointed RXTE observations with the Proportional
Counter Array (PCA; Jahoda et al.~1996) and the High-Energy X-ray Timing Experiment
(HEXTE; Rothschild et al.~1998) instruments.  The first 14 RXTE observations were part
of a guest observer program with results reported by Cui et al.~(1999).  They found
that during the initial X-ray rise (0.7--2.4~Crab at 2--10~keV), the source exhibited
very strong quasiperiodic X-ray oscillations (QPOs) in the range 0.08--8~Hz.  The
spectral and timing analysis of 60 additional RXTE observations, reported in Sobczak
et al.~(1999a) and Remillard et al.~(1999a), revealed the presence of canonical
outburst states characteristic of black hole X-ray novae (see Tanaka \& Lewin 1995)
and X-ray QPOs at a few Hz and $\sim200$~Hz.  XTE~J1550--564 is one of only a few
Galactic black hole candidate known to exhibit QPOs above 50~Hz (Remillard et
al.~1999a; Homan, Wijnands, \& van der Klis 1999); the others are 4U~1630--47
(Remillard \& Morgan 1999) and the microquasars GRS~1915$+$105 (Morgan, Remillard, \&
Greiner 1997) and GRO~J1655--40 (Remillard et al.~1999b).  

The X-ray light curve of XTE~J1550--564 from the ASM aboard RXTE is shown in Figure~1.
The outburst exhibits a `double-peaked' profile, with the first half generally
dominated by power-law emission, and the second half generally dominated by emission
from the accretion disk (see \S3).  The double-peaked profile of the outburst is
different from the outbursts of classical X-ray novae like A0620--00 (see Chen,
Shrader, \& Livio 1997), but is similar to the outburst behavior of the microquasar
GRO~J1655--40 (Sobczak et al.~1999b).  

The optical (Orosz, Bailyn, \& Jain 1998) and radio (Campbell-Wilson et al.~1998)
counterparts of XTE~J1550--564 were identified shortly after the source was
discovered.  The presence of an optical counterpart, with $B\sim22$~mag in quiescence
(Jain et al.~1999), is especially important since this will allow radial velocity
studies of the companion star during quiescence that could confirm the black hole
nature of the primary.  

Herein we present spectral results for 209 X-ray observations spanning the entire 250
days of the 1998-99 outburst of XTE~J1550--564.  These observations include the rising
phase observations with results first reported by Cui et al.~(1999) (RXTE program
P30188-06), all of our RXTE guest observer program (P30191), and all of the public
observations of this source (P30435 \& P40401).  The spectral analysis of PCA
observations 15-75 (see Table~1) was first reported by Sobczak et al.~(1999a).  A
timing study based on those same RXTE observations is presented in Remillard et
al.~(1999a) and observations of the optical counterpart are presented in Jain et
al.~(1999).  Low frequency QPOs (0.08--18~Hz) were observed during 74 of the 209
observations reported in the present paper.  The frequencies, amplitudes, and
coherence factors (Q) of these QPOs can be found in Table~1 of Sobczak et al.~(2000).

\section{Observations and Analysis}

We present spectral results for 209 observations of XTE~J1550--564 (see Fig.~1)
obtained using the PCA and HEXTE instruments onboard RXTE.  The PCA consists of five
xenon-filled detector units (PCUs) with a total effective area of
$\sim$~6200~cm$^{-2}$ at 5~keV.  The PCA is sensitive in the range 2--60~keV, the
energy resolution is $\sim$17\% at 5~keV, and the time resolution capability is
1~$\mu$sec.  The HEXTE consists of two clusters of 4 NaI/CsI phoswich scintillation
detectors, each with an effective area of 800~cm$^2$, covering the energy range 15
to 250 keV with an energy resolution of 9~keV at 60~keV.  A journal of the PCA/HEXTE
observations of XTE~J1550--564 is given in Table~1, including exposure times, count
rates, etc.  

The PCA and HEXTE pulse height spectra were not fit simultaneously because of
uncertainty in the cross-calibration of the instruments.  This uncertainty is apparent
when fitting the spectrum of the Crab.  Fitting the Crab spectrum (e.g.~1997
December~15 and 1999 March~24) to a power-law using either PCA or HEXTE data alone
yields $\chi^2_{\nu} \sim 1$ for each detector; however, fitting PCA and HEXTE data
simultaneously, even while floating the relative normalization of the detectors,
results in $\chi^2_{\nu} \sim 4$.  For this reason, we present separate fits to the
PCA and HEXTE spectra.

\subsection{PCA}

The PCA data were taken in the ``Standard 2'' format, which consists of 129 channel
spectra accumulated for each PCU every 16~s.  The data span PCA gain epochs~3~\&~4
(epoch~4, which covers observations 170--209, began when the PCA gain was lowered on
1999 March~22 at 16:30 UT).   The epoch~3 response matrix for each PCU was obtained
from the 1998~January distribution of response files, and the response matrix for
epoch~4 was obtained from the RXTE GOF website and dated 1999 March~31.  We tested the
latest epoch~4 response matrices released with FTOOLS v5.0 and found that there are no
significant changes in the fits.  The pulse height spectrum from each PCU was fit over
the energy range 2.5--20~keV for epoch~3, using a systematic error in the count rates
of 1\%.  For epoch~4, the pulse height spectrum from each PCU was fit over the energy
range 3--20~keV, using a systematic error in the count rates of 2\%.  The lower limit
of the energy range was raised to 3~keV for epoch~4 because the sensitivity of the
detectors at low energy decreased when the gain was lowered, and a systematic error of
2\% was adopted in order to maintain $\chi^2_{\nu} \sim 1$ for the Crab nebula.  All
of the spectra were corrected for background using the standard bright source
background models appropriate for each epoch.  We began using the epoch~4 faint source
background model beginning on 1999 April~27, when the source count rate dropped below
40~c/s/PCU.  Only PCUs~0~\&~1 were used for the spectral fitting reported here and the
spectra from both PCUs were fit simultaneously using XSPEC (Sobczak et al.~1999a,b).  

The PCA spectral data were fit to the widely used model consisting of a multicolor
blackbody accretion disk plus power-law (Tanaka \& Lewin 1995; Mitsuda et al.~1984;
Makishima et al.~1986).  The fits were significantly improved by including a smeared
Fe absorption edge near 8~keV (Ebisawa et al.~1994; Inoue 1991) and a weak Fe emission
line; the best-fit line had a central energy around 6.5 keV, a width held fixed at 1.2
keV (FWHM), and an equivalent width $\lesssim100$~eV.  Interstellar absorption was
modeled using the Wisconsin cross-sections (Morrison \& McCammon 1983).  All of the
observations were first fit with a floating hydrogen column density ($N_H$), which was
generally in the range 1.5--2.5~$\times 10^{22}$~cm$^{-2}$ (Jain et al.~1999 estimated
$N_H=0.9 \times 10^{22}$~cm$^{-2}$ from optical reddening).  The fits were insensitive
to $N_H$ differences in this small range, and in the final analysis presented here,
$N_H$ was fixed at $2.0 \times 10^{22}$~cm$^{-2}$.  There are a total of eight free
parameters: the apparent temperature ($T_{in}$) and radius ($R_{in}$) of the inner
accretion disk, the power-law photon index ($\Gamma$) and normalization ($K$), the
edge energy and optical depth ($\tau_{Fe}$) of the Fe absorption feature, and the
central energy and normalization of the Fe emission line.

The addition of the Fe emission and absorption components is motivated in
Figures~2a~\&~2b, which show the ratio of a typical spectrum to the best fit model
without and with the Fe emission and absorption.  The addition of the Fe emission and
absorption components reduces the $\chi^2_{\nu}$ from 7.9 to 0.9 in this example.  
The Fe line and edge energies also agree with the relation in Figure~17 of Nagase
(1989) and indicate variations in the ionization state of Fe during the outburst.

The fitted temperature and radius of the inner accretion disk presented here ($T_{in}$
and $R_{in}$) are actually the color temperature and radius of the inner disk, which
may be affected by spectral hardening due to electron scattering (Shakura \& Sunyaev
1973; Shimura \& Takahara 1995).  The physical interpretation of these parameters
remains uncertain and is discussed below.  The reader should also note that the inner
disk radius is obtained from the normalization of the multicolor disk model, $R_{in}
(\cos~i)^{1/2} / (D/6~kpc)$, which is a function of the distance $D$ and inclination
$i$ of the system.  We use $i=0$ and $D=6$~kpc for XTE~J1550--564, but the actual
distance and inclination are unknown.  Six representative spectra are shown in
Figures~3a--3f.  The model parameters and component fluxes (see Tables~2 \& 3) are
plotted in Figures~4a--4f.  All uncertainties are given at the $1\sigma$ confidence
level.  Unless otherwise noted, the spectral parameters discussed in this paper are
those derived using the PCA as opposed to the HEXTE spectra.  

\subsection{HEXTE}

The standard HEXTE reduction software was used for the extraction of the HEXTE archive
mode data.  The HEXTE modules were alternatingly pointed every 32~s at source and
background positions, allowing background subtraction with high sensitivity to time
variations in the particle flux at different positions in the spacecraft orbit. 
Clusters A and B were fit simultaneously and the normalization of each cluster was
allowed to float independently, since there is a small systematic difference between
the normalizations of the two clusters.  We used the HEXTE response matrices released
1997 March~20.  Only the data above 20~keV was used because of uncertainty in the
response at lower energies.  

The source was not detected in the HEXTE during all observations.  For those
observations in which the source was detected, the HEXTE spectra were fit to a
power-law model from 20~keV to the maximum energy at which the source was detected,
which ranged from 50 to 200~keV.  In a number of cases we found that the HEXTE spectra
could not be adequately fit using a pure power-law model in the observed energy range
(Fig.~5a--5d).  In these instances, we used a power-law with a high energy cutoff of
the form (cf. Grove et al.~1998): 
\begin{mathletters} 
\begin{eqnarray} N(E) & = & K
E^{-\Gamma}~~~~~for~E \leq E_{cut} \\ & = & K E^{-\Gamma} \exp \left( \left(E_{cut} -
E \right)/E_{fold} \right)~~~~~for~E \geq E_{cut} 
\end{eqnarray} 
\end{mathletters} 

The addition of the high energy cutoff improved the value of $\chi^2_{\nu}$ from 4.9
to 0.7 for the observation on 1998~Sept.~7 (Fig.~5) and gave similarly dramatic
improvements for many other observations.  A three parameter cutoff model (the
equivalent of $E_{cut}=0$) can also fit the data in some cases.  Similarly, the
`comptt' model in XSPEC~v.10 (Titarchuk 1994) can be used to fit most of these data,
but it tends to cutoff more rapidly than the data at high energy.  These data can also
be adequately fit by a model including a pure power-law component plus a broad
gaussian or reflection component.  The presence of a reflection component is possible,
given the inclusion of Fe absorption when fitting the PCA spectra (see above).  If a
reflection component is present, the fit parameters indicate that it would contribute
as much as half of the 2--20~keV or 2--100~keV flux for the 1998 Sept.~7 observation. 
Observations at higher energies are necessary to determine whether the high energy
cutoff seen in the 20--200~keV range is due to a physical cutoff in the power-law
component or an underlying reflection feature.  

The HEXTE model parameters are given in Table~4.  The parameters of the high energy
cutoff are given only for those cases where the addition of the cutoff improved the
value of $\chi^2_{\nu}$ by more than 30\%.  The value of $\chi^2_{\nu}$ for the pure
power-law model is also given for comparison in these cases.  Upper limits for the
20--100~keV HEXTE flux were determined by assuming a fixed power-law photon index of
2.5.  The reader should also note that the normalization of the HEXTE instrument is
systematically $\sim20$--30\% lower than the PCA; we have not adjusted our spectral
fits of the HEXTE data to correct for this discrepancy.

\section{Spectral Results}

Six representative spectra are shown in Figures~3a--3f.  These spectra illustrate the
range of X-ray spectra for XTE~J1550--564, in which (a) a strong power-law with high
energy cutoff dominates a warm disk (rising phase), (b) an intense power-law component
dominates a hot disk (very high state flare), (c,d) the disk dominates a weak
power-law (high/soft state), (e) a strong power-law dominates a warm disk, and (f) a
weak power-law dominates a cool disk.  

As discussed in the introduction, the outburst of XTE~J1550--564 can be divided into
two halves, which we now discuss in turn.  We define the first half as extending from
the start of the outburst on MJD~51062 (1998 September~6) to approximately MJD~51150
(1998 December~3).  During the initial rise (MJD~51063--51072), the spectra are
dominated by the power-law component with a photon index that gradually steepens from
$\Gamma = 1.5$ to 2.5 (Fig.~4c, Table~2), and the source displays strong 0.08--8~Hz
QPOs (Cui et al.~1999).  A high energy cutoff is observed in the HEXTE spectrum during
the initial rise with $E_{fold} \sim$~70--150 (Fig.~3a, Table~4).  Following the
initial rise, the spectra from MJD~51074 to 51115 remain dominated by the power-law
component (Fig.~3b) which has photon index $\Gamma\sim$~2.4--2.9 (Fig.~4c).  The
source displays strong 3--13~Hz QPOs during this time, whenever the power-law
contributes more than 60\% of the observed X-ray flux (Remillard et al.~1999a; Sobczak
et al.~1999a).  This behavior is consistent with the {\it very high state} of Black
Hole X-ray Novae (BHXN) (see Tanaka \& Lewin (1995) and references therein for details
on the spectral states of BHXN).  A high energy cutoff in the power-law tail is
observed during a number of very high state observations (Table~4).  The peak
luminosity (bolometric disk luminosity plus 2--100~keV power-law luminosity) during
the flare on MJD~51075 is $L=1.2\times10^{39}(D/6kpc)^2$ erg~s$^{-1}$, which
corresponds to the Eddington luminosity for $M=9.6M_{\odot}$ at 6~kpc (Sobczak et
al.~1999a).  

After MJD~51115 (1998 October~29), the source fades rapidly, the power-law component
decays and hardens ($\Gamma = 2.0-2.4$), and the disk component begins to dominate the
spectrum (see Fig.~3c).  The source generally shows little temporal variability during
this time (Remillard et al.~1999a; Sobczak et al.~1999a).  We identify this behavior
with the {\it high/soft state}.  However, during this time the source occasionally
exhibits QPOs at $\sim$~5~Hz and power density spectra that have properties
intermediate between the very high and high/soft states or the high/soft and low/hard
states (Remillard et al.~1999a; Sobczak et al.~1999a).  The {\it low/hard state} was
not observed and the intensity increased dramatically after MJD~51150, marking the
start of the second half of the outburst.  

During the second half of the outburst (MJD~51150--51230), most of the observed
spectra are dominated by the disk component (Fig.~3d), which contributes $>85$\% of
the 2--20~keV flux (Fig.~4e), and no QPOs are observed.  Figure~4c shows that $\Gamma$
increased sharply at the onset of the second half of the outburst from $\Gamma\sim$~2
to 4.  These features are typical of the {\it high/soft state} of BHXN.  This
dichotomy between the power-law-dominated first half and disk-dominated second half of
the outburst is shared by GRO~J1655--40, and cannot be easily explained by the
standard disk instability model.  

Also during the second half of the outburst, there are a few instances lasting two or
three days when the power-law hardens from $\Gamma \sim4$ to 2.5 (Fig.~4c).  During
one of these instances (MJD~51201), the power-law flux increases by almost a factor of
two and the source is also detected in the HEXTE (see Table~4).  After MJD~51230 (1999
February~21), the power-law hardens considerably (Fig.~4c) and there is an intense
power-law flare, which begins on MJD~51241 (1999 March~4; see Fig.~4d), accompanied by
a sharp decline in the disk flux (Fig.4e).  QPOs from 5--18~Hz also reappear during
this power-law flare.  A sample spectrum from this flare is shown in Figure~3e.  A
high energy cutoff in the power-law component is marginally detected in two
HEXTE observations during the second half of the outburst (Table~4).  The total flux
decreases steadily as the power-law flare fades after MJD~51260 (1999 March~23; see
Fig.~4f).  The spectrum of one of the last few observations resembles the low/hard
state and is shown in Figure~3f.  

A comparison of the 2-12~keV (soft) PCA flux and the 20--100~keV (hard) HEXTE flux is
shown in Figure~6.  During the first few observations, the hard flux exceeds or is
approximately equal to the soft flux.  Following the initial rise, during the very
high state in the first half of the outburst, the soft flux exceeds the hard flux by
almost an order of magnitude.  During the high/soft state in the second half of the
outburst, the source is usually undetectable in the HEXTE, and the upper limits on the
hard flux show that the soft flux exceeds the hard flux by more than 2.5 orders of
magnitude.  This strong dominance of the soft flux over the hard flux is
characteristic of the high/soft state.  The hard flux increases again during the
power-law flare and subsequent decline that mark the end of the outburst.  Similar
variations in the relative strength of the soft and hard flux between outburst states
were also observed for the BHXN Nova Muscae~1991 (Esin, McClintock, \& Narayan 1997)
and are a good means of differentiating the very high and high/soft states.  

From Figure~4b, it appears that the inner radius of the disk does not appear to be
constant throughout the outburst cycle.  From Figure~4b and Table~2, we see that the
intense 6.8~Crab flare on MJD~51075 (1999 September~19) is accompanied by a dramatic
decrease of the inner disk radius from 33 to 2~km over one day.  Similar behavior was
observed for GRO~J1655--40 during its 1996-97 outburst: The observed inner disk radius
decreased by almost a factor of four during periods of increased power-law emission in
the very high state, and it was generally larger in the high/soft state (Sobczak et
al.~1999b).  Below we discuss both the problems and possible interpretations that are
relevant to our spectral results, focussing on the observed variation of the inner
disk radius.

\section{Discussion}

\subsection{The Inner Disk Radius}

The physical radius of the inner disk may vary in XTE~J1550--564 and GRO~J1655--40 --
by as much as a factor of 16 in one day in the case of XTE~J1550--564.  Another
possibility, however, is that the apparent decrease of the inner disk radius observed
during intense flares is caused by the failure of the multicolor disk/power-law model
at these times.  This failure may be caused by one or both of the following effects:
(1) {\it Variations in spectral hardening, which occurs because electron scattering as
opposed to free-free absorption dominates the opacity in the inner disk} (Shakura \&
Sunyaev 1973).  This causes the observed (color) temperature of the disk to appear
higher than the effective temperature, which decreases the normalization of the
multicolor disk model from which the radius is inferred (Ebisawa et al.~1994; Shimura
\& Takahara 1995).  The spectral hardening correction is likely not to be constant in
extreme situations, such as the 6.8~Crab flare, and any increase in spectral hardening
will appear as a decrease in the observed inner disk radius.  (2) {\it The Compton
upscattering of soft disk photons, which is likely the origin of the power-law
component in BHXN.}  In this case, an increase in power-law emission naturally implies
an increase in the Comptonization of soft disk photons.  Therefore, the measured
normalization of the multicolor disk (from which the radius is derived) represents
only a fraction of the intrinsic X-ray emission from the disk because the photons that
are upscattered to produce the power-law component are missing.  This causes the
inferred radius to decrease (Zhang et al.~1999), although a corresponding increase in
the mass accretion rate is necessary to explain the associated increase in the
apparent disk temperature.  Thus, intense flares may cause an apparent decrease in the
radius of the inner disk due to increased spectral hardening and/or Compton
upscattering of soft disk photons, while the actual physical radius may remain fairly
constant.  

The above interpretation is bolstered by the constancy of the inner disk radius of
XTE~J1550--564 when the power-law is weak: The observed inner disk radius remains
approximately constant at $\sim40$~km (assuming $i=0\arcdeg$, $D=6$~kpc) over 120 days
from MJD~51120--51240 when the power-law contributes less than 20\% of the 2--20~keV
flux, even though the disk flux and total flux vary by an order of magnitude.  The
observed inner disk radius in GRO~J1655--40 also remains approximately constant over
more than 150 days during the high state, when the power-law contributes less than
10\% of the flux (Sobczak et al.~1999b).  Similar behavior has been observed for
several other BHXN and Galactic black hole candidates, where the observed stable inner
disk radius was plausibly identified with the last stable orbit (Tanaka \& Lewin
1995).  If we hypothesize that the stable value of the radius for XTE~J1550--564 is a
reasonable measure of the radius of the last stable orbit ($6~r_g$ where $r_g =
GM/c^2$), then a drop by a factor of 20 (observed between the peak flare and the
high/soft state) would imply a decrease from $6~r_g$ to $0.3~r_g$, which is well
within the event horizon of the black hole.  This result is independent of the assumed
distance and inclination.  We thus conclude that the small inner disk radius observed
during the 6.8~Crab flare is unphysical and that the inner disk radius during the most
intense power-law activity cannot be reliably determined from the multicolor disk
model.  This interpretation is supported by the work of Merloni, Fabian, \& Ross
(1999) who used a self-consistent model for radiative transfer and vertical
temperature structure in a Shakura-Sunyaev disk and found that (1) the multicolor disk
model systematically underestimates the inner disk radius when most of the
gravitational energy is dissipated in the corona and (2) the multicolor disk model
gives stable, acceptable results for high accretion rates ($\ga 10$\% of the Eddington
accretion rate) and/or when a lower fraction of the gravitational energy is dissipated
in the corona.  

In contrast to the apparent decrease of the inner disk radius observed during the
6.8~Crab flare, the radius appears to increase at times near the beginning and end of
the outburst cycle (Fig.~4b).  It is not clear whether these observed increases of the
inner disk radius are physical or not; in the following we discuss two possible
causes.  

The Advection-Dominated Accretion Flow (ADAF) model predicts changes in the inner disk
radius during the initial rise and the final decline of the outburst.  According to
the ADAF model, the inner edge of the disk should move inward (from hundreds of 
gravitational radii) to the last stable orbit during the initial rise, and it should
move outward again at the end of the outburst during the transition from the high/soft
state to the low/hard state (Esin, McClintock, \& Narayan 1997).  However, the ADAF
model predicts that the radius should move inward on a viscous timescale of several
days (Hameury et al.~1998) during the initial rise; instead, about 30 days elapsed
between the initial rise and the time when the observed inner disk radius reached an
approximately constant value (Fig.~4b).  On the other hand, the observed increase of
the inner disk radius near the end of the outburst from MJD~51279--51293 is in better
agreement with the predictions of the ADAF model.  The critical mass accretion rate at
which the soft-to-hard transition takes place in the ADAF model is approximately
$\dot{m}_{crit} \sim 1.3 \alpha^2$ (in Eddington units), where $\alpha$ is the
standard viscosity coefficient (Esin et al.~1997).  The ADAF model predicts that the
inner edge of the disk should begin to move outward from the last stable orbit when
$\dot{m} \lesssim \dot{m}_{crit}$.  We can estimate the mass accretion rate
($\dot{m}$) of XTE~J1550--564 in Eddington units by assuming that the 6.8~Crab flare
corresponds to the Eddington accretion rate and scaling the inferred accretion rate by
the total observed flux.  The ratio of the total observed flux on MJD~51279, when the
observed radius begins to increase, to the total flux during the 6.8~Crab flare
corresponds to a mass accretion rate $\dot{m} \sim 0.02$ (Table~3).  The mass
accretion rate at this time corresponds to the critical accretion rate for $\alpha =
0.12$, which is smaller than the value $\alpha = 0.25$ used by Esin et al.~(1997) to
model the outburst of Nova Muscae 1991, but is still in reasonable agreement with the
predictions of the ADAF model.  

It is also possible that the observed increase of the inner disk radius at early times
is not physical, but rather due to the flattening of the radial temperature profile
from the $T \sim r^{3/4}$ assumed in the multicolor disk model.  This effect would
cause the radius inferred from the multicolor disk model to increase.  The flattening
of the radial temperature profile in the disk could be caused by irradiation of the
disk during the initial rise and following the intense 6.8~Crab flare.

\subsection{Black Hole Mass}

We can estimate the mass of the black hole in XTE~J1550-564 by equating the observed
inner disk radius with the last stable orbit predicted by general relativity.  Merloni
et al.~(1999) find that the multicolor disk model, with appropriate corrections for
spectral hardening and relativistic effects, gives stable, acceptable results for high
accretion rates ($\ga 10$\% of the Eddington accretion rate) and/or when a lower
fraction of the gravitational energy is dissipated in the corona.  In this case, the
actual inner disk radius ($R_{in}$) can be determined from the {\it stable} observed
(color) radius ($r_{in}$) during the high/soft state according to the formula:
\begin{equation} 
R_{in} = \eta g(i) f^2 r_{in}, 
\end{equation} 
where $\eta$ is the ratio of the inner disk radius to the radius at which the disk
emissivity peaks, $g(i)$ is the correction for relativistic effects as a function of
inclination, $f$ is the correction for constant spectral hardening ($f=1.7$; Shimura
\& Takahara~1995), and $r_{in} = 40$~km~$(D/6~kpc) (\cos i)^{-1/2}$ for XTE~J1550--564
(see above).  Both $\eta$ and $g(i)$ are functions of black hole angular momentum. 
Assuming a Schwarzschild black hole with $0\arcdeg \leq i \leq 70\arcdeg$, for which
$\eta=0.63$ and $g(i) \sim 0.9$ (Zhang, Cui, \& Chen 1997; Sobczak et al.~1999a),
eq.~(2) becomes $R_{in}=66$~km~$(D/6~kpc) (\cos i)^{-1/2}$.  Equating $R_{in}$ with
the last stable orbit for a Schwarzschild black hole ($6GM/c^2$), we find:
\begin{equation}
M_{BH} = 7.4~M_{\odot}(D/6~kpc) (\cos i)^{-1/2}.
\end{equation}
For example, for $i=60\arcdeg$ and $D=6$~kpc, $M_{BH}= 10~M_{\odot}$.

\section{Conclusion}

We have analyzed RXTE data acquired during observations of the X-ray nova
XTE~J1550--564.  Satisfactory fits to all the PCA data were obtained with a model
consisting of a multicolor disk, a power-law, and Fe emission and absorption
components.  XTE~J1550--564 is observed in the very high and high/soft canonical
outburst states of BHXN.  We distinguished these two states based on the relative flux
contribution of the disk and power-law spectral components, the value of the power-law
photon index, the presence or absence of QPOs, and the relative strength of the
2--12~keV and 20--~100~keV fluxes.

The source exhibited an intense (6.8~Crab) flare on MJD~51075, during which the inner
disk radius appears to have decreased dramatically from 33 to 2~km (for $i=0$ and
$D=6$~kpc).  However, the apparent decrease of the inner disk radius observed during
periods of increased power-law emission may be caused by the failure of the multicolor
disk model; the actual physical radius of the inner disk may remain fairly constant. 
This interpretation is supported by the fact that the observed inner disk radius
remains approximately constant over 120 days from MJD~51120--51240, when the power-law
is weak, even though the disk flux and total flux vary by an order of magnitude.  The
mass of the black hole inferred by equating the approximately constant inner disk
radius observed in the high/soft state with the last stable orbit for a Schwarzschild
black hole (see \S~4.2) is $M_{BH} = 7.4~M_{\odot}(D/6~kpc) (\cos i)^{-1/2}$.  

The outburst of XTE~J1550--564 has many features in common with the most recent
outburst of the microquasar GRO~J1655--40 (Sobczak et al.1999b).  During the first
half of their outbursts, the X-ray spectra of both sources are dominated by the
power-law component and both the observed inner disk radius and flux exhibit extreme
variability.  Following their initial outbursts, the flux from both sources declined,
only to be followed by a second outburst.  During the second half of their outbursts,
the X-ray spectra of both sources are primarily disk-dominated with an approximately
constant inner disk radius and slowly varying intensity.  Neither of these sources can
be described as a `canonical' BHXN.

\acknowledgements This work was supported, in part, by NASA grants NAG5-3680 and
NAS5-30612.  Partial support for J.M. and G.S. was provided by the Smithsonian
Institution Scholarly Studies Program.  W.C. would like to thank Shuang Nan Zhang and
Wan Chen for extensive discussions on spectral modeling and interpretation of the
results.

\newpage 

\begin{figure}
\figurenum{1}
\plotone{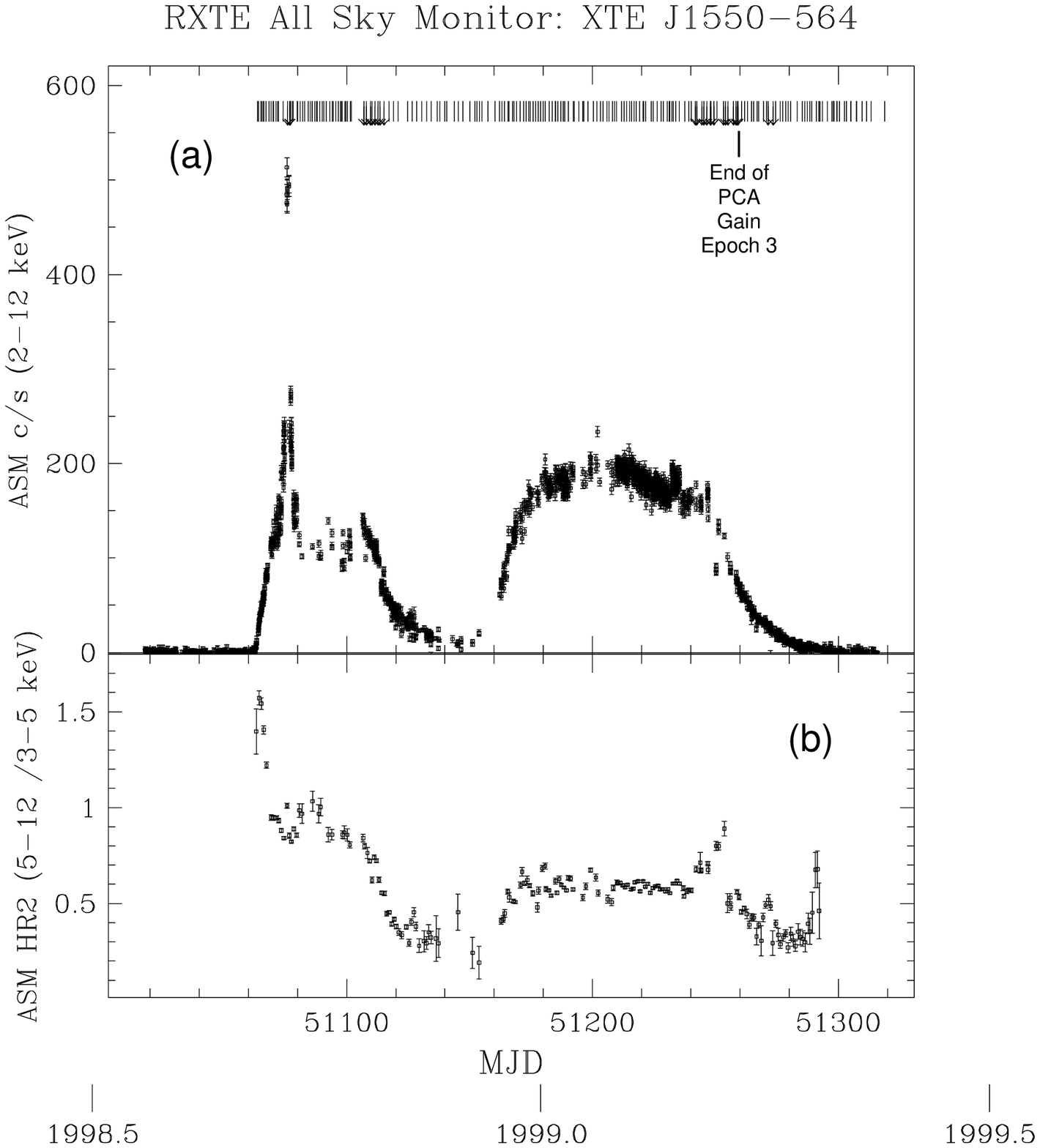}
\caption{\singlespace (Upper Panel) The 2--12~keV ASM lightcurve and (Lower Panel) the
ratio of the ASM count rates (5--12~keV)/(3--5~keV) for XTE~J1550-564.  The small,
solid vertical lines in the top panel indicate the times of pointed RXTE observations;
the downward arrows indicate the observations during which high frequency 161--283~Hz
QPOs are present.  }
\end{figure}

\begin{figure}
\figurenum{2}
\plotone{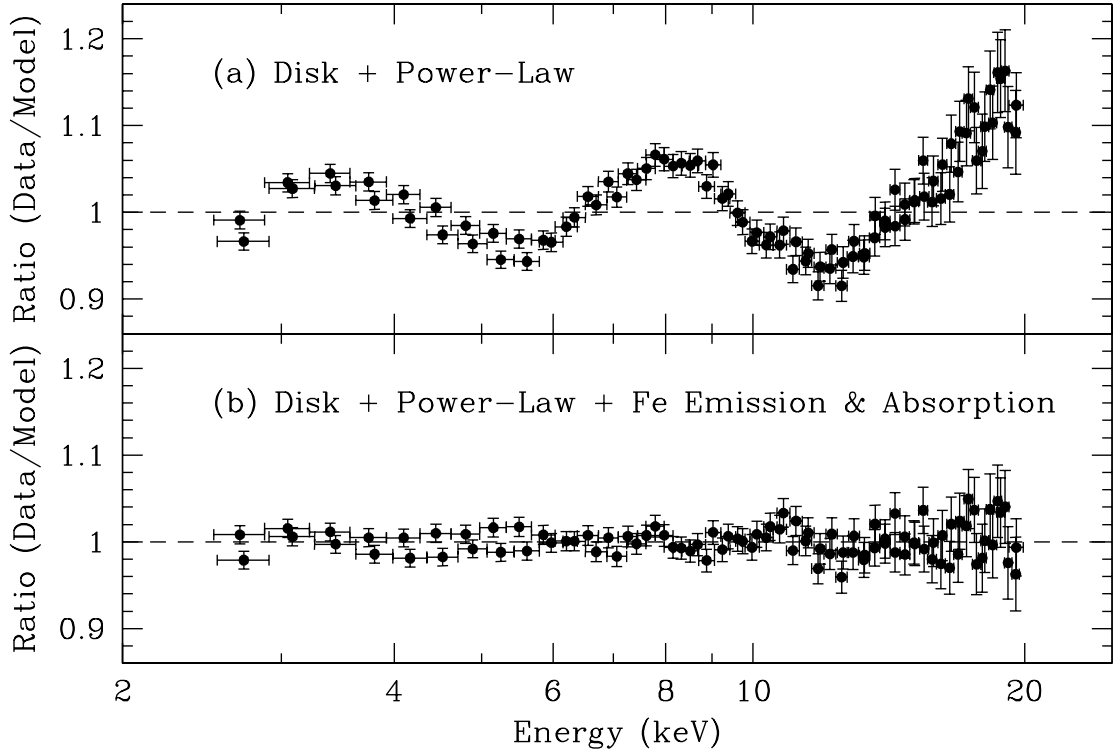}
\caption{\singlespace The ratio data/model for (a) the best fit multicolor disk plus
power-law model and (b) the multicolor disk plus power-law plus Fe emission \&
absorption model for a representative high/soft state spectrum (MJD~51126, 1998
Nov.~9).  The addition of the Fe emission \& absorption components improves the
$\chi^2_{\nu}$ from (a) 7.9 to (b) 0.9 in this example.  }
\end{figure}

\begin{figure}
\figurenum{3}
\plotone{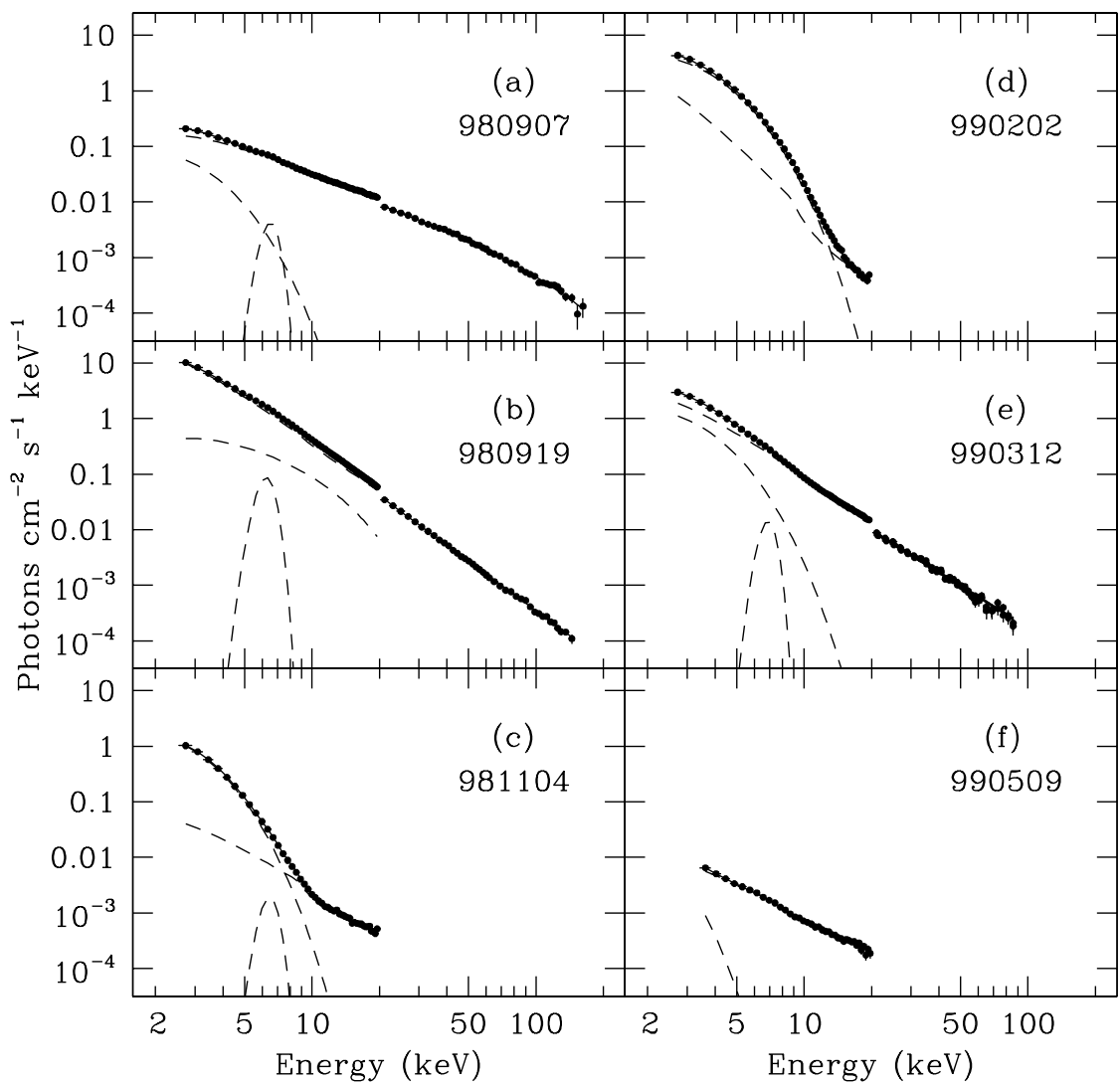}
\caption{\singlespace Sample PCA spectra from (a) the rising phase on MJD~51063 (1998
Sept.~7), (b) the flare on MJD~51075 (1998 Sept.~19), (c,d) the high/soft state on
MJD~51121 \& 51211 (1998 Nov.~4 \& 1999 Feb.~2), (e) the power-law flare during the
decline of the outburst (MJD~51249; 1999 March~12), and (f) one of the last
observations on MJD~51307 (1999 May~9) during gain epoch~4.  The individual components
of the model are also shown (dashed lines).  Although error bars are plotted for all
the data, they are usually not large enough to be visible.  }
\end{figure}

\begin{figure}
\figurenum{4}
\plotone{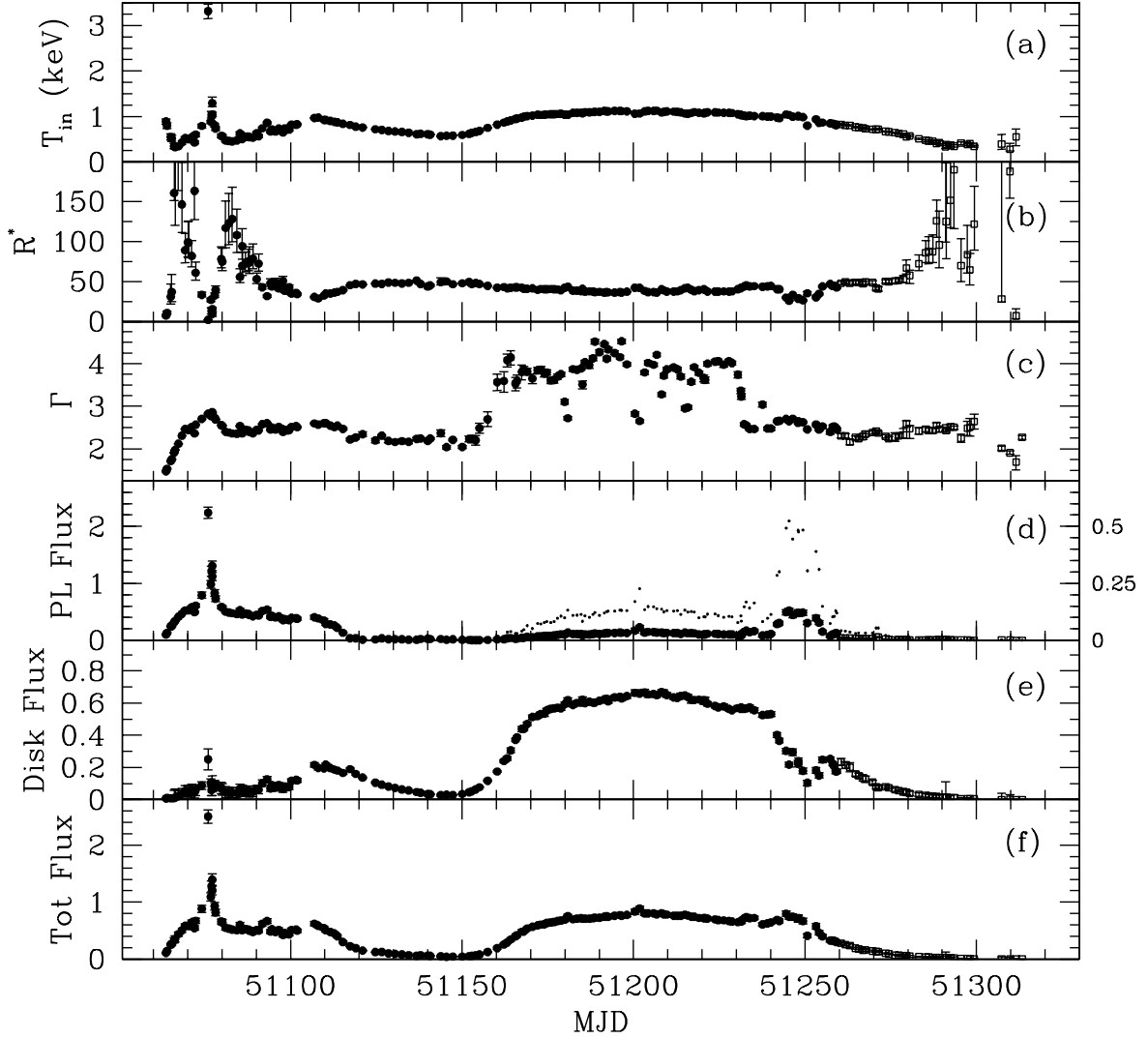}
\caption{\singlespace Spectral parameters and fluxes for PCA observations of
XTE~J1550--564.  See the text for details on the spectral models and fitting.  The
quantities plotted here are (a) the color temperature of the accretion disk $T_{in}$
in keV, (b) the inner disk radius $R^*=R_{in}(\cos~i)^{1/2}/(D/6~kpc)$ in km, where
$i$ is the inclination angle and $D$ is the distance to the source in kpc, (c) the
power-law photon index $\Gamma$, the unabsorbed 2--20~keV flux in units of
$10^{-7}$~erg~s$^{-1}$~cm$^{-2}$ for (d) the power-law, (e) the disk, and (f) the
total.  Data points from gain epoch~4, beginning on MJD~51260 (MJD = JD--2,400,000.5),
are plotted using an open square.  When error bars are not visible, it is because they
are comparable to or smaller than the plotting symbol.  The dots plotted without error
bars in (d) correspond to the right axis and are shown to highlight the behavior of
the faint power-law component during the second half of the outburst.}
\end{figure}

\begin{figure}
\figurenum{5}
\plotone{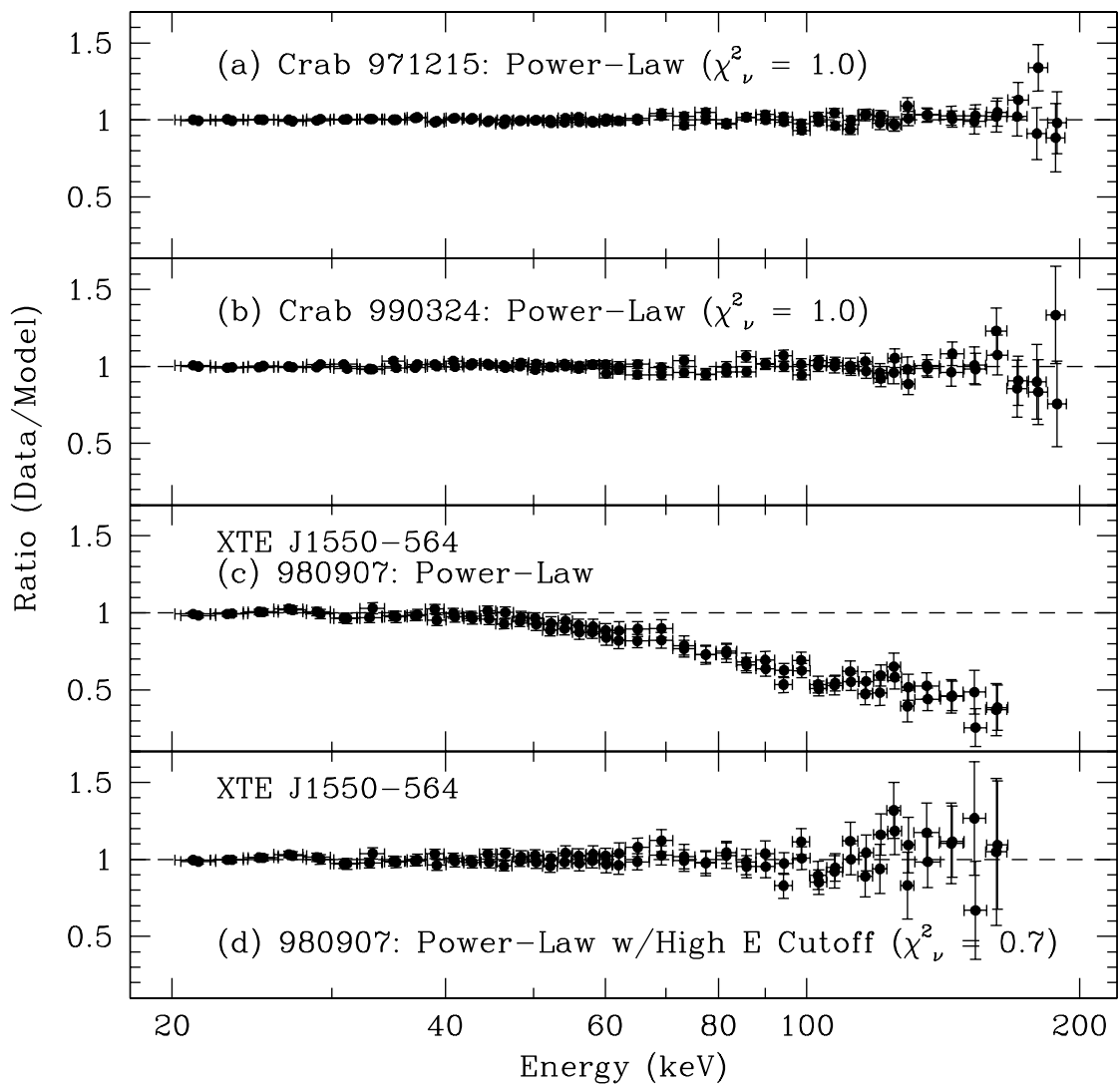}
\caption{\singlespace The ratio data/model for (a) power-law fit to the Crab on 1997
December~15, (b) later power-law fit to the Crab on 1999 March~24, (c) power-law fit
to the first observation of XTE~J1550--564 on 1998 September~7, and (d) the same
observation of XTE~J1550--564 fit to a model consisting of a power-law with a high
energy cutoff (see Eq.~1).  }
\end{figure}

\begin{figure}
\figurenum{6}
\plotone{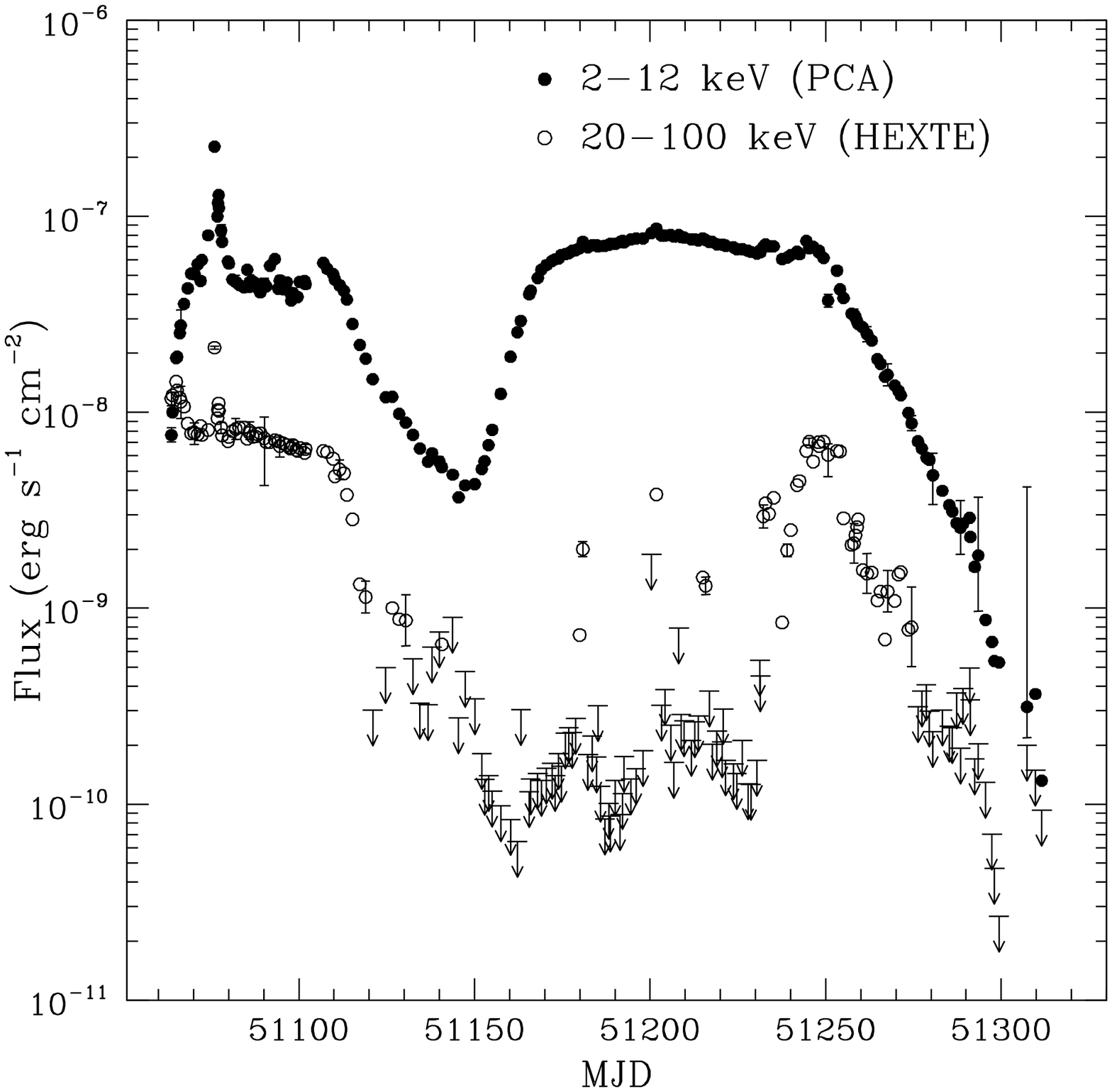}
\caption{\singlespace Plot of the 2--12~keV PCA flux and the 20--100~keV HEXTE flux vs. 
time (MJD = JD--2,400,000.5).  Representative error bars are plotted for every 5th
point.  Upper limits for the HEXTE data are plotted at the $3\sigma$ level of
confidence (recall that the normalization of the HEXTE instrument is systematically
$\sim20$--30\% lower than the PCA).  }
\end{figure}


\newpage
\begin{references}
\reference{}Campbell-Wilson, D., McIntyre, V., Hunstead, R., \& Green, A. 1998,
\iaucirc~7010
\reference{}Chen, W., Shrader, C. R., \& Livio, M. 1997, \apj, 491, 312
\reference{}Cui, W., Zhang, S. N., Chen, W., \& Morgan, E. H. 1999, \apj, 512, L43
\reference{}Ebisawa, K., Ogawa, M., Aoki, T., Dotani, T., Takizawa, M., Tanaka, Y.,
Yoshida, K., Miyamoto, S., Iga, S., Hayashida, K., Kitamoto, S., \&
Terada, K. 1994, \pasj, 46, 375
\reference{}Esin, A. A., McClintock, J. E., \& Narayan, R. 1997, \apj, 489, 865
\reference{}Grove, J. E., Johnson, W. N., Kroeger, R. A., McNaron-Brown, K., Skibo, J.
G., Hulburt, E. O., \& Philips, B. F. 1998, \apj, 500, 899
\reference{}Hameury, J.-M., Lasota, J.-P., McClintock, J. E., \& Narayan, R. 1998,
\apj, 489, 234
\reference{}Homan, J., Wijnands, R., \& van der Klis, M. 1999, \iaucirc~7121
\reference{}Inoue, H. 1991, in Frontiers of X-ray Astronomy, ed. Y. Tanaka \& K. Koyama
(Tokyo:  Universal Academy Press), 291
\reference{}Jahoda, K., Swank, J. H., Giles, A. B., Stark, M. J., Strohmayer, T.,
Zhang, W., \& Morgan, E. H. 1996, Proc. SPIE 2808, ``EUV and Gamma Ray Instumentation
for Astronomy'' VII, 59
\reference{}Jain, R., Bailyn, C. D., Orosz, J. A., Remillard R. A., \& 
McClintock, J. E. 1999, \apj, 517, L131
\reference{}Levine, A. M., Bradt, H., Cui, W., Jernigan, J. G., Morgan, E. H.,
Remillard, R., Shirey, R. E., \& Smith, D. A. 1996, \apjl, 469, 33
\reference{}Makishima, K., Maejima, Y., Mitsuda, K., Bradt, H. V., Remillard, R.
A., Tuohy, I. R., Hoshi, R., \& Nakagawa, M. 1986, \apj, 308, 635
\reference{}Merloni, A., Fabian, A. C., \& Ross, R. R. 1999, \mnras, in press 
(astro-ph/9911457)
\reference{}Mitsuda, K., Inoue, H. Koyama, K., Makishima, K., Matsuoka, M.,
Ogawara, Y., Shibazaki, N., Suzuki, K., Tanaka, Y., \& Hirano, T. 1984, PASJ, 36, 741
\reference{}Morgan, E. H., Remillard, R. A. \& Greiner, J. 1997, \apj, 482, 993
\reference{}Morrison, R. \& McCammon, D. 1983, \apj, 270, 119
\reference{}Nagase, F. 1989, \pasj, 41, 1
\reference{}Orosz, J., Bailyn, C., \& Jain, R. 1998, \iaucirc~7009
\reference{}Remillard, R. A., McClintock, J. E., Sobczak, G. J., Bailyn, C. D., Orosz,
J. A., Morgan, E. H., \& Levine, A. M. 1999a, \apj, 517, L127
\reference{}Remillard, R. A., Morgan, E. H., McClintock, J. E., Bailyn, C. D., \& 
Orosz, J. A. 1999b, \apj, 522, 397
\reference{}Remillard, R. A. \& Morgan, E. H. 1999, AAS Meeting, 195, 3702 
\reference{}Rothschild, R. E., Blanco, P. R., Gruber, D. E., Heindl, W. A., MacDonald,
D. R., Marsden, D. C., Pelling, M. R., Wayne, L. R., \& Hink, P. L. 1998, \apj, 496, 538
\reference{}Shakura, N. I. \& Sunyaev, R. A. 1973, \aap, 24, 337
\reference{}Shimura, T. \& Takahara, F. 1995, \apj, 445, 780
\reference{}Smith, D. A. \& RXTE/ASM teams 1998, \iaucirc~7008
\reference{}Sobczak, G. J., McClintock, J. E., Remillard, R. A., Levine, A. M., 
Morgan, E. H., Bailyn, C. D., \& Orosz, J. A. 1999a, \apj, 517, L121
\reference{}Sobczak, G. J., McClintock, J. E., Remillard, R. A., Bailyn, C. D., \&
Orosz, J. A. 1999b, \apj, 520, 776
\reference{}Sobczak, G. J., McClintock, J. E., Remillard, R. A., Cui, W., Levine, A.
M., Morgan, E. H., Orosz, J. A., \&  Bailyn, C. D. 2000, \apj, 531, 537
\reference{}Tanaka, Y. \& Lewin, W. H. G. 1995, in X-ray Binaries, ed. W. H. G. Lewin,
J. van Paradijs, \& E. P. J. van den Heuvel (Cambridge:  Cambridge Univ. Press), p. 126
\reference{}Titarchuk, L. 1994, \apj, 434, 570
\reference{}Zhang, S. N., Cui, W., \& Chen, W. 1997, \apj, 482, L155
\reference{}Zhang, S. N., et al.~1999, in preparation
\end{references}
\end{document}